# Free iterative and iteration $K$-semialgebras

Z. Ésik[*]
Dept. of Computer Science
University of Szeged
Hungary

W. Kuich[†]
Inst. für Diskrete Mathematik
TU Wien
Austria

**Abstract**

We consider algebras of rational power series over a finite alphabet $\Sigma$ with coefficients in a commutative semiring $K$ and characterize them as the free algebras in various classes of algebraic structures.

## 1 Introduction

One of the most important algebraic structures that emerged in Theoretical Computer Science is the algebra $\mathbf{Reg}_\Sigma$ of regular languages over the finite alphabet $\Sigma$, equipped with the regular operations $+$, $\cdot$ and Kleene star $^*$. The importance of this structure is underlined by the fact that for any finite alphabet $\Sigma$, $\mathbf{Reg}_\Sigma$ may be characterized in an abstract way as the free algebra over $\Sigma$ in the variety generated by all algebras of languages equipped with the regular operations, or by the algebras of binary relations, where Kleene star is interpreted as the formation of the reflexive-transitive closure, or by continuous idempotent semirings, where the star operation provides least fixed point solutions to linear equations of the sort $x = ax+1$, cf. [9, 10, 21].

Regular languages over $\Sigma$ equipped with the regular operations form an iteration semiring [4, 13]. It was shown by Krob in [24] (see also [4, 12]) that for any finite alphabet $\Sigma$, $\mathbf{Reg}_\Sigma$ is freely generated by $\Sigma$ in the subvariety of iteration semirings characterized by the single equation $1^* = 1$.

Regular languages have been generalized to rational power series in [30]. For recent excellent treatments of rational series the reader is referred to [3, 28]. When $K$ is a semiring and $\Sigma$ is a finite alphabet, the semiring of rational power series $K^{\mathrm{rat}}\langle\!\langle \Sigma^*\rangle\!\rangle$ may be equipped with a partial star operation, defined on the proper series. Equipped with this star operation, $K^{\mathrm{rat}}\langle\!\langle \Sigma^*\rangle\!\rangle$ is both a partial iteration semiring and an iterative semiring [6]. When $K$ is an iteration semiring

[*]Partially supported by the project TÁMOP-4.2.1/B-09/1/KONV-2010-0005 "Creating the Center of Excellence at the University of Szeged", supported by the European Union and co-financed by the European Regional Fund, the Austrian-Hungarian Action Foundation, grant 77öu9, the National Foundation of Hungary for Scientific Research, grant no. K 75249, and the ESF project AUTOMATHA.
[†]Partially supported by the Austrian-Hungarian Action Foundation, grant 77öu9.

(and thus has a total star operation) [4, 13], then $K^{\text{rat}}\langle\!\langle\Sigma^*\rangle\!\rangle$ is also an iteration semiring. We may equip $K^{\text{rat}}\langle\!\langle\Sigma^*\rangle\!\rangle$ with a natural $K$-action turning $K^{\text{rat}}\langle\!\langle\Sigma^*\rangle\!\rangle$ to a $K$-semimodule. When $K$ is a commutative semiring, $K^{\text{rat}}\langle\!\langle\Sigma^*\rangle\!\rangle$ is a partial iteration $K$-semialgebra and an iterative $K$-semialgebra, or an iteration $K$-semialgebra when $K$ is a commutative iteration semiring, as defined in Sections 3, 8 and 9, respectively. In this paper, our aim is to characterize the semirings $K^{\text{rat}}\langle\!\langle\Sigma^*\rangle\!\rangle$ as the free algebras in the class of iterative $K$-semialgebras and the class of (partial) iteration $K$-semialgebras, for commutative semirings and for commutative iteration semirings $K$, satisfying some additional conditions. Our results provide both a generalization of Salomaa's axiomatization [27] of regular languages, based on the unique fixed point rule, and a generalization of Krob's axiomatization [24] of regular languages, and of recent axiomatizations [5] of rational power series with coefficients in the the semiring $\mathbb{N}$ of natural numbers, and the semiring $\mathbb{N}_\infty$ of natural numbers endowed with a top element $\infty$.

## 2 Semirings, semimodules and semialgebras

Recall from [18, 19, 20, 25] that a *semiring* $S = (S, +, \cdot, 0, 1)$ consists of a commutative monoid $(S, +, 0)$ and a monoid $(S, \cdot, 1)$ such that multiplication (or product) $\cdot$ distributes over addition (or sum) $+$, and moreover $x \cdot 0 = 0 = 0 \cdot x$ for all $x \in S$. A semiring $S$ is called *commutative* if $xy = yx$ for all $x, y \in S$, and *idempotent* if $x + x = x$ for all $x \in S$. Morphisms of semirings preserve the sum and product operations and the constants $0, 1$. (When writing expressions, we will follow the standard convention that multiplication has higher precedence than addition.) Examples of semirings include all fields and rings, all bounded distributive lattices including the 2-element lattice $\mathbb{B} = \{0, 1\}$, called the *Boolean semiring*, the semiring $\mathbb{N}$ of natural numbers, the semiring $\mathbb{N}_\infty$ obtained from $\mathbb{N}$ by adding a point $\infty$ at infinity, so that $n + \infty = \infty$ and $m\infty = \infty$ for all $n, m$ such that $m \neq 0$, and the *tropical semiring* $\mathbf{T} = (\mathbb{N} \cup \{\infty\}, \min, +, \infty, 0)$ which has the same domain as the semiring $\mathbb{N}_\infty$, where sum is the minimum operation and product is ordinary addition with $n + \infty = \infty + n = \infty$, for all $n \in \mathbb{N} \cup \{\infty\}$, and whose constants are $\infty$ and $0$. In order to avoid trivial situations, we will only consider nontrivial semirings in which $0 \neq 1$. When $S$ is a semiring, so is the collection $S^{n \times n}$ of all $n \times n$ ($n \geq 1$) matrices over $S$ with the usual operations and constants. We identify any matrix in $S^{1 \times n}$ with the corresponding row vector, and any matrix in $S^{n \times 1}$ with the corresponding column vector.

When $S$ is a semiring, an *$S$-semimodule* is a commutative monoid $V = (V, +, 0)$ equipped with a (left) $S$-action $S \times V \to V$, $(s, v) \mapsto sv$ subject to the usual laws:

$$\begin{aligned}
(s + s')a &= sa + s'a \\
s(a + b) &= sa + sb \\
(ss')a &= s(s'a) \\
1a &= a \\
0a &= 0 \\
s0 &= 0
\end{aligned}$$

for all $s, s' \in S$ and $a, b \in V$. Morphisms of $S$-semimodules preserve the additive structure and the $S$-action. Note that each semiring $S$ is an $S$-semimodule whose $S$-action $S \times S \to S$ is the product operation of the semiring. More generally, for each $n, p \geq 1$, the set $S^{n \times p}$ of all $n \times p$ matrices over $S$ equipped with the pointwise sum operation, the zero matrix as constant 0, and the pointwise action is an $S$-semimodule. Note also that any semiring is naturally an $\mathbb{N}$-semimodule, and any idempotent semiring is a $\mathbb{B}$-semimodule.

Suppose now that $K$ is a commutative semiring. An (associative) $K$-*semialgebra* [20] is a semiring $A$ which is a $K$-semimodule satisfying

$$k(ab) \quad = \quad (ka)b \quad = \quad a(kb) \tag{1}$$

for all $k \in K$ and $a, b \in A$.

Since $K$ is commutative, the left $K$-action determines a right $K$-action by $ak := ka$, for all $k \in K$ and $a \in A$. In a $K$-semialgebra, we will usually denote the multiplicative identity by $e$. Morphisms of $K$-semialgebras preserve the semiring operations and the $K$-action. Any semiring may be viewed as an $\mathbb{N}$-semialgebra, and any idempotent semiring is a $\mathbb{B}$-semialgebra.

**Remark 2.1** *We defined $K$-semialgebras for commutative semirings only, since if (1) holds, then so does $(kk')(ab) = (ka)(k'b) = (k'k)(ab)$, for all $a, b \in A$ and $k, k' \in K$, so that if the $K$-action is faithful, then $K$ is commutative. Nevertheless some of the facts in the paper will hold for non-commutative semirings.*

## 2.1 Formal series

Suppose that $S$ is a semiring and $\Sigma$ is a finite alphabet. A *series* [3, 11, 18, 20, 25, 29] over $\Sigma$ with coefficients in $S$ is a formal sum

$$s = \sum_{w \in \Sigma^*} (s, w)w$$

where $(s, w) \in S$ for all $w \in \Sigma^*$. Here, $\Sigma^*$ denotes the free monoid of all words over $\Sigma$ including the empty word $\epsilon$. Note that each series $s$ can be viewed as a function $\Sigma^* \to S$. We let $S\langle\!\langle \Sigma^* \rangle\!\rangle$ denote the set of all such series.

The sum and product operations on series $s, s' \in S\langle\!\langle \Sigma^* \rangle\!\rangle$ are defined by

$$\begin{aligned}(s + s', w) &= (s, w) + (s', w) \\ (ss', w) &= \sum_{uv=w} (s, u)(s', v), \quad w \in \Sigma^*.\end{aligned}$$

As usual, we identify each element $s \in S$ with the series such that the coefficient of $\epsilon$ is $s$ and all other coefficients are 0. This defines the constant series 0 and 1. We also identify each word $w \in \Sigma^*$ with the series that maps $w$ to 1 and all other words to 0.

It is well-known that equipped with these operations and constants, $S\langle\!\langle \Sigma^* \rangle\!\rangle$ is a semiring. We may also define an $S$-action by $(sr, w) = s(r, w)$ for all $s \in S$ and $r \in S\langle\!\langle \Sigma^* \rangle\!\rangle$, so that $S\langle\!\langle \Sigma^* \rangle\!\rangle$ becomes an $S$-semimodule. When $S$ is commutative, $S\langle\!\langle \Sigma^* \rangle\!\rangle$ is an $S$-semialgebra.

The *support* of a series $s \in S\langle\!\langle \Sigma^* \rangle\!\rangle$ is defined as the set $\mathrm{supp}(s) = \{w \in \Sigma^* : (s, w) \neq 0\}$. We call a series $s \in S\langle\!\langle \Sigma^* \rangle\!\rangle$ *proper* if $(s, \epsilon) = 0$, i.e., when $\epsilon \notin \mathrm{supp}(s)$. Each language $L \subseteq \Sigma^*$ is the support of its *characteristic series* $s \in \mathbb{B}\langle\!\langle \Sigma^* \rangle\!\rangle$ defined by $(s, w) := 1$ if $w \in L$ and $(s, w) := 0$, otherwise. The semiring $\mathbb{B}\langle\!\langle \Sigma^* \rangle\!\rangle$ is isomorphic to the semiring $P(\Sigma^*)$ of all subsets of $\Sigma^*$ equipped with set union as sum and concatenation as product. The constants 0 and 1 are the empty language $\emptyset$ and the set $\{\epsilon\}$.

We call a series $s \in S\langle\!\langle \Sigma^* \rangle\!\rangle$ a *polynomial* if $\mathrm{supp}(s)$ is finite. The collection $S\langle \Sigma^* \rangle$ of all polynomials is a subsemiring and an $S$-subsemimodule of $S\langle\!\langle \Sigma^* \rangle\!\rangle$. When $S$ is commutative, $S\langle \Sigma^* \rangle$ is an $S$-semialgebra. The following fact is standard.

**Theorem 2.2** *Suppose that $K$ is a commutative semiring and $\Sigma$ is a finite alphabet. Then $K\langle \Sigma^* \rangle$ is freely generated by $\Sigma$ in the class of all $K$-semialgebras.*

The content of this result is that for any $K$-semialgebra $A$ and function $h : \Sigma \to A$, there is a unique $K$-semialgebra morphism $h^\sharp$ extending $h$. To define $h^\sharp$, first extend $h$ to a monoid morphism $\Sigma^* \to A$, and then further extend it to a linear map.

## 3 Iterative semialgebras and partial Conway semialgebras

An ideal of a semiring $S$ is a set $I \subseteq S$ such that $0 \in I$, $I + I \subseteq I$, $SI \cup IS \subseteq I$. Note that if $I \subseteq S$ is an ideal and $1 \in I$, then $S = I$. Recall from [6] that an *iterative semiring* is a semiring $S$ with a distinguished ideal $I$ such that for any $a \in I$ and $b \in S$, the equation $x = ax + b$ has a *unique solution* in $S$. In a *symmetric iterative semiring*, equations $x = xa + b$ with $a \in I$ and $b \in S$ also have unique solutions. Morphisms of (symmetric) iterative semirings preserve the distinguished ideal.

In the rest of this section, we let $K$ denote a commutative semiring. Suppose that $A$ is a $K$-semialgebra. An ideal $I \subseteq A$ is called a $K$-*ideal* if $KI \subseteq I$.

**Definition 3.1** *Suppose that $A$ is a $K$-semialgebra with a distinguished $K$-ideal $I$. When $A$ is also a (symmetric) iterative semiring then we call $A$ a* (symmetric) iterative $K$-*semialgebra. Morphisms of (symmetric) iterative $K$-semialgebras are $K$-semialgebra morphisms that preserve the distinguished $K$-ideal.*

Note that if $A$ is an iterative $K$-semialgebra with distinguished $K$-ideal $I$, then $e \notin I$ since otherwise $A$ would be trivial.

**Example 3.2** *When $K = \mathbb{N}$ and $A$ is a $K$-semialgebra, a $K$-ideal is just an ideal. Thus, every (symmetric) iterative semiring is a (symmetric) iterative $\mathbb{N}$-semialgebra. Similarly, every idempotent (symmetric) iterative semiring is a (symmetric) iterative $\mathbb{B}$-semialgebra.*

**Example 3.3** *If $\Sigma$ is any finite alphabet, then $K\langle\!\langle \Sigma^* \rangle\!\rangle$ is a symmetric iterative $K$-semialgebra with the collection of all proper series as distinguished $K$-ideal. See [3, 25].*

**Example 3.4** *Suppose that $K$ is a commutative ring and that $A$ is a $K$-semialgebra with distinguished $K$-ideal $I$ such that the multiplicative inverse $(e - a)^{-1}$ exists for each $a \in I$. Since $K$ is a ring, so is $A$. Moreover, for each $a \in I$ and $b \in A$, $(e - a)^{-1}b$ is the unique solution of the equation $x = ax + b$, and $b(e - a)^{-1}$ is the unique solution of the equation $x = xa + b$. Thus, $(A, I)$ is a symmetric iterative $K$-semialgebra.*

Suppose that $A$ is an iterative $K$-semialgebra with distinguished $K$-ideal $I$. Then $A$ may be equipped with a *star operation* $^* : I \to A$ such that for each $a \in I$, $a^*$ is the unique solution of the equation $x = ax + e$. It follows that for any $a \in I$ and $b \in A$, $a^*b$ is the unique solution of the equation $x = ax + b$, since $aa^*b + b = (aa^* + e)b = a^*b$.

For the following result, see Proposition 5.1 in [6].

**Theorem 3.5** *The following hold in an iterative semiring $S$ with distinguished ideal $I$:*

$$\begin{align}
(a+b)^* &= (a^*b)^*a^*, \quad a, b \in I \tag{2} \\
(ab)^* &= 1 + a(ba)^*b, \quad a \in I \text{ or } b \in I. \tag{3}
\end{align}$$

*A similar fact holds in any iterative $K$-semialgebra $A$ with distinguished $K$-ideal $I$. Any morphism of iterative semirings or iterative $K$-semialgebras preserves the star operation.*

In [6], a *partial Conway semiring* is defined as a semiring $S$ with a distinguished ideal $I$ and a star operation $^* : I \to S$ such that (2) and (3) hold. Morphisms of partial Conway semirings preserve the distinguished ideal and the star operation.

**Remark 3.6** *A commutative semiring $K$ equipped with a distinguished ideal $I$ and a star operation $^* : I \to K$ is a partial Conway semiring iff (2) holds and $aa^* + 1 = a^*$ for all $a \in I$.*

**Definition 3.7** *Suppose that $A$ is a $K$-semialgebra, $I$ is a distinguished $K$-ideal of $A$, and suppose that we are given a star operation $^* : I \to A$. We call $(A, I, ^*)$ a partial Conway $K$-semialgebra if it is a partial Conway semiring. A morphism of partial Conway $K$-semialgebras is a $K$-semialgebra morphism that preserves the distinguished $K$-ideal and the star operation.*

Below, when $I$ and $^*$ are understood, we will just write $A$ for a partial Conway $K$-semialgebra $(A, I, ^*)$. Later we will also define Conway $K$-semialgebras that are partial Conway $K$-semialgebras $(A, I, ^*)$ with $A = I$. But in that case, we will also require that there is a star operation on $K$ which is compatible with the star operation on $A$. It is clear that every partial Conway semiring is a partial Conway $\mathbb{N}$-semialgebra, and every idempotent partial Conway semiring is a partial Conway $\mathbb{B}$-semialgebra.

**Corollary 3.8** *Any iterative $K$-semialgebra is uniquely a partial Conway $K$-semialgebra. Any morphism of iterative $K$-semialgebras is a partial Conway $K$-semialgebra morphism.*

In particular, when $K$ is a commutative semiring and $\Sigma$ is a finite alphabet, then $K\langle\!\langle \Sigma^* \rangle\!\rangle$ with distinguished ideal the collection of all proper series is a partial Conway $K$-semialgebra. In any partial Conway $K$-semialgebra $(A, I, ^*)$, we define $a^+ := aa^* = a^*a$, for all $a \in I$. We call this operation the *plus operation*. Note that $a^+ \in I$ for all $a \in I$, and moreover, the plus operation in turn determines the star operation since $a^* = a^+ + e$, for all $a \in I$.

Perhaps the most important result for partial Conway semirings and partial Conway $K$-semialgebras is a general version of Kleene's theorem. In order to state this result, we extend the star operation to matrices by the usual matrix star formula. Since every partial Conway semiring is a partial Conway $\mathbb{N}$-semialgebra, the definition below also applies to partial Conway semirings.

Let $M \in I^{n \times n}$, where $(A, I, ^*)$ is a partial Conway $K$-semialgebra. When $n = 1$, $M = (a)$, for some $a \in I$, and we define $M^* = (a^*)$. Suppose now that $M = \begin{pmatrix} X & Y \\ U & V \end{pmatrix}$ where $X \in I^{k \times k}$, $Y \in I^{k \times 1}$, $U \in I^{1 \times k}$ and $V \in I^{1 \times 1}$, and suppose that we have already defined the star of any matrix over $I$ of size $m \times m$, where $m < n$. Then we define

$$\begin{pmatrix} X & Y \\ U & V \end{pmatrix}^* = \begin{pmatrix} \alpha & \beta \\ \gamma & \delta \end{pmatrix} \quad (4)$$

where

$$\begin{aligned} \alpha &= (X + YV^*U)^* & \beta &= \alpha YV^* \\ \gamma &= \delta UX^* & \delta &= (V + UX^*Y)^*. \end{aligned}$$

It can be seen by induction that the star of a matrix is well-defined. For example, $V^*$ is well-defined by the induction hypothesis, and since all entries of $(X + YV^*U)$ are in $I$, $(X + YV^*U)^*$ is also well-defined by the induction hypothesis. See [4, 6, 8] for more details. For further use we note that when $M \in I^{n \times n}$, then $M^+ = MM^* \in I^{n \times n}$.

The next result corresponds to Theorem 5.1 in [6]. For Theorem 3.10, see [4, 6, 8].

**Theorem 3.9** *Suppose that $A$ is an iterative $K$-semialgebra with distinguished $K$-ideal $I$ and hence a partial Conway $K$-semialgebra in a canonical way. Then for each $n \geq 1$, $(A^{n \times n}, I^{n \times n})$ is also an iterative $K$-semialgebra. Moreover, for each $M \in I^{n \times n}$ and $N \in A^{n \times p}$, the equation $X = MX + N$ has as unique solution in the variable $X$ ranging over $A^{n \times p}$ the matrix $M^*N$, where $M^*$ can be computed by the matrix star formula (4). When $A$ is a symmetric iterative $K$-semialgebra with distinguished $K$-ideal $I$, then for each $n \geq 1$, $(A^{n \times n}, I^{n \times n})$ is also a symmetric iterative $K$-semialgebra, and when $M \in I^{n \times n}$ and $N \in A^{p \times n}$, $NM^*$ is the unique solution of the fixed point equation $X = XM + N$ in the variable $X$ ranging over $A^{p \times n}$.*

**Theorem 3.10** *If $(A, I, ^*)$ is a partial Conway $K$-semialgebra, then so is the matrix $K$-semialgebra $(A^{n \times n}, I^{n \times n}, ^*)$, for each $n \geq 1$, where star is defined by the matrix star formula (4). Moreover, the matrix star formula holds for all partitions of $M \in I^{n \times n}$ into matrices $X, Y, U, V$, where $X$ and $V$ are square matrices.*

## 4 Automata in partial Conway $K$-semialgebras

In this section we state a variant of a general Kleene-type result for automata in Conway $K$-semialgebras, where $K$ is a commutative semiring.

**Definition 4.1** *Suppose that $\mathcal{A} = (A, I, ^*)$ is a partial Conway $K$-semialgebra and $\Sigma \subseteq I$. Let $K\Sigma$ denote the $K$-semimodule generated by $\Sigma$ in $A$. A $K$-automaton over $\Sigma$ in $A$ is a triplet*

$$\mathcal{A} = (\alpha, M, \beta)$$

*where $\alpha \in K^{1 \times n}$, $M \in (K\Sigma)^{n \times n}$, $\beta \in K^{n \times 1}$ are the* initial vector, *the* transition matrix *and the* final vector *of $\mathcal{A}$, respectively. The integer $n \geq 1$ is called the* dimension *of $\mathcal{A}$. The* behavior *of $\mathcal{A}$ is*

$$|\mathcal{A}| := \alpha M^* \beta = \alpha e \beta + \alpha M^+ \beta = \alpha \beta e + \alpha M^+ \beta \in A.$$

*Two automata are called* equivalent *if they have the same behavior.*

(Here, $\alpha M^* \beta$ is defined similarly to the matrix product using the left action (and the induced right action) of $K$ on $A$.)

**Definition 4.2** *Suppose that $A = (A, I, {}^*)$ is a partial Conway $K$-semialgebra and $\Sigma \subseteq I$. We say that $a \in A$ is $K$-recognizable over $\Sigma$ in $A$ if there is a $K$-automaton $\mathcal{A}$ over $\Sigma$ in $A$ with $|\mathcal{A}| = a$. We let $\mathrm{Rec}_A(\Sigma)$ denote the set of all $K$-recognizable elements over $\Sigma$.*

**Definition 4.3** *Suppose that $A = (A, I, {}^*)$ is a partial Conway $K$-semialgebra and $\Sigma \subseteq I$. We say that $a \in A$ is $K$-rational over $\Sigma$ in $A$ if $a$ can be written in the form $a = ke + b$, where $k \in K$ and $b$ can be generated from $\Sigma \cup \{0\}$ by the sum, product, $K$-action and plus operations. We let $\mathrm{Rat}_A(\Sigma)$ denote the collection of all $K$-rational elements over $\Sigma$.*

The following Kleene-type result follows from Theorem 6.1 in [6].

**Theorem 4.4** *For any partial Conway $K$-semialgebra $A = (A, I, {}^*)$ and any $\Sigma \subseteq I$,*

$$\mathrm{Rec}_A(\Sigma) = \mathrm{Rat}_A(\Sigma).$$

*Moreover, if either $e \in I$ or for all $k \in K$ and $a \in I$, $ke + a \in I$ implies that $ke = 0$, then $\mathrm{Rat}_A(\Sigma)$ with distinguished $K$-ideal $I' = \mathrm{Rat}_A(\Sigma) \cap I$ and ${}^*$ restricted to $I'$ is a partial Conway $K$-semialgebra, the least partial Conway $K$-subsemialgbera of $A$ containing $\Sigma$.*

For any finite alphabet $\Sigma$, let $K^{\mathrm{rec}}\langle\!\langle \Sigma^* \rangle\!\rangle := \mathrm{Rec}_{K\langle\!\langle \Sigma^* \rangle\!\rangle}(\Sigma)$ and $K^{\mathrm{rat}}\langle\!\langle \Sigma^* \rangle\!\rangle := \mathrm{Rat}_{K\langle\!\langle \Sigma^* \rangle\!\rangle}(\Sigma)$. A series in $K^{\mathrm{rec}}\langle\!\langle \Sigma^* \rangle\!\rangle$ is called *$K$-recognizable*, and a series in $K^{\mathrm{rat}}\langle\!\langle \Sigma^* \rangle\!\rangle$ is called *$K$-rational*, cf. [3, 28, 29].

**Corollary 4.5** *For any finite alphabet $\Sigma$, $K^{\mathrm{rec}}\langle\!\langle \Sigma^* \rangle\!\rangle = K^{\mathrm{rat}}\langle\!\langle \Sigma^* \rangle\!\rangle$ is a (symmetric) iterative $K$-semialgebra and thus a partial Conway $K$-semialgebra with the set of proper rational series as distinguished $K$-ideal. Moreover, $K^{\mathrm{rat}}\langle\!\langle \Sigma^* \rangle\!\rangle$ is the least partial Conway $K$-subsemialgebra of $K\langle\!\langle \Sigma^* \rangle\!\rangle$ containing $\Sigma$.*

## 5 Simulations of automata

In this section, we assume that $K$ is a fixed commutative semiring. The notion of simulation was introduced in [4, 14] in order to relate equivalent automata. It can be traced back to Schützenberger's result on the minimization of weighted automata over fields, cf. [3]. Simulations over the Boolean semiring are implicit in [21]. Simulation is called "conjugacy" in [1]. The notion of (functional) simulation is closely related to the notion of bisimulation, see [4].

**Definition 5.1** *Suppose that $A = (A, I, {}^*)$ is a partial Conway $K$-semialgebra and $\Sigma \subseteq I$. Let $\mathcal{A} = (\alpha, M, \beta)$ and $\mathcal{B} = (\gamma, N, \delta)$ be automata over $\Sigma$ in $A$ of dimension $m$ and $n$, respectively. We say that a matrix $X \in K^{m \times n}$ is a simulation $\mathcal{A} \to \mathcal{B}$, denoted $\mathcal{A} \to^X \mathcal{B}$, if*

$$\alpha X = \gamma, \quad MX = XN, \quad \beta = X\delta.$$

*A* functional simulation *is a simulation $X$ such that each row of $X$ contains exactly one occurrence of $1$, and all other entries are $0$. A* dual functional simulation *is a simulation $X$ satisfying the same condition for columns. A* diagonal simulation *is a simulation by a diagonal matrix.*

*We say that $\mathcal{A}$ and $\mathcal{B}$ are* simulation equivalent *if there is a finite sequence of automata $\mathcal{C}_i$ together with matrices $X_i$ of appropriate size, for $i = 0, \ldots, k-1$, such that $\mathcal{C}_0 = \mathcal{A}$, $\mathcal{C}_k = \mathcal{B}$, and for each $0 \leq i < k$, either $\mathcal{C}_{i+1} \to^{X_i} \mathcal{C}_i$ or $\mathcal{C}_i \to^{X_i} \mathcal{C}_{i+1}$.*

Note that if $\mathcal{A} \to^X \mathcal{B}$ and $\mathcal{B} \to^Y \mathcal{C}$, then $\mathcal{A} \to^{XY} \mathcal{C}$, so that we may require that in the above definition of simulation equivalence, the simulations are "alternating".

**Lemma 5.2** *Suppose that $A$ is an iterative $K$-semialgebra with distinguished $K$-ideal $I$ and let $\Sigma \subseteq I$. Suppose that $\mathcal{A}$ and $\mathcal{B}$ are $K$-automata over $\Sigma$ in $A$ such that there is a simulation $\mathcal{A} \to \mathcal{B}$. Then $\mathcal{A}$ and $\mathcal{B}$ are equivalent.*

Proof. Let $\mathcal{A} = (\alpha, M, \beta)$ and $\mathcal{B} = (\gamma, N, \delta)$, and suppose that $X$ is a simulation $\mathcal{A} \to \mathcal{B}$. We show that $M^*X = XN^*$. Indeed, since

$$M(XN^*) + X = XNN^* + X = XN^*,$$

we have $M^*X = XN^*$, by Theorem 3.9 and Theorem 3.10. Thus,

$$|\mathcal{A}| = \alpha M^* \beta = \alpha M^* X \delta = \alpha X N^* \delta = \gamma N^* \delta = |\mathcal{B}|. \quad \square$$

**Corollary 5.3** *If $A$ is an iterative $K$-semialgebra with distinguished $K$-ideal $I$ and if the $K$-automata $\mathcal{A}$ and $\mathcal{B}$ over $\Sigma \subseteq I$ are simulation equivalent, then $\mathcal{A}$ and $\mathcal{B}$ are equivalent.*

# 6 Proper semirings

In this section, $K$ again denotes a commutative semiring.

Corollary 5.3 applies to the (symmetric) iterative $K$-semialgebras $K\langle\!\langle \Sigma^* \rangle\!\rangle$ and $K^{\mathrm{rat}}\langle\!\langle \Sigma^* \rangle\!\rangle$, where $\Sigma$ is any finite alphabet. In the next definition, by an automaton in $K\langle\!\langle \Sigma^* \rangle\!\rangle$ we shall mean an automaton over $\Sigma$ in the iterative $K$-semialgebra $K\langle\!\langle \Sigma^* \rangle\!\rangle$ whose distinguished $K$-ideal is the set of all proper series.

**Definition 6.1** *Following [17], we call $K$* proper *if for any finite alphabet $\Sigma$ and $K$-automata $\mathcal{A}$ and $\mathcal{B}$ in $K\langle\!\langle \Sigma^* \rangle\!\rangle$, if $\mathcal{A}$ and $\mathcal{B}$ are equivalent, then they are simulation equivalent.*

In the next result, we give a sufficient condition of properness. We call a semiring $S$ *Noetherian* if every subsemimodule of a finitely generated $S$-semimodule is finitely generated. For example, all finite semirings, all fields and all finitely generated commutative rings are Noetherian, cf. [3]. The semiring $\mathbb{N}$ is not Noetherian, since $\mathbb{N}^2$ is a finitely generated $\mathbb{N}$-semimodule and the subsemimodule generated by the pairs $(1,1), (2,1), \ldots, (n,1), \ldots$ is not finitely generated. The following fact was proved in [17] using techniques developed in [1, 26].

**Theorem 6.2** *If every finitely generated subsemiring of the commutative semiring $K$ embeds in a Noetherian subsemiring, then $K$ is proper. In particular, if $K$ is Noetherian, then $K$ is proper.*

The fact that $\mathbb{B}$ and all finite commutative semirings are proper was shown in [4, 21] and [14], respectively.

In the sequel, we will also make use of a refinement of the notion of proper semirings. Recall that a (multiplicative) *unit* of a semiring is simply an invertible element. An *invertible diagonal simulation* is a simulation by a diagonal matrix whose diagonal entries are units.

**Definition 6.3** *We say that $K$ is* strongly proper *if for any finite alphabet $\Sigma$ and $K$-automata $\mathcal{A}$ and $\mathcal{B}$ in $K\langle\!\langle \Sigma^* \rangle\!\rangle$, if $\mathcal{A}$ and $\mathcal{B}$ are equivalent, then $\mathcal{A}$ and $\mathcal{B}$ can be connected by a finite chain of functional, dual functional, and invertible diagonal simulations.*

Clearly, every strongly proper semiring is proper.

Following [1], we call a semiring $S$ *equisubstractive* if for all $x, y, z, u \in S$, if $x + y = z + u$ then there exist $a, b, c, d \in S$ with $x = a + b$, $y = c + d$, $z = a + c$, $u = b + d$. Moreover, we say that a semiring $S$ is *additively generated by its units* if any element of $S$ is a (possibly empty) finite sum of units. For example, $\mathbb{N}$, $\mathbb{Z}$ and all fields are additively generated by their units. Clearly, if $S$ is additively generated by its units, then so is any quotient of $S$. In particular, for each $k \geq 2$, the semiring $\mathbb{N}_k$ obtained from $\mathbb{N}$ by collapsing all natural numbers $\geq k - 1$ is additively generated by its single unit 1. (When $k = 2$, then $\mathbb{N}_k$ is just the Boolean semiring.)

The following important fact is taken from [1].

**Theorem 6.4** *If $K$ is equisubstractive and additively generated by its units, then $K$ is proper iff $K$ is strongly proper.*

## 6.1 Atomicity

Atomicity is a stronger form of equisubstarctiveness. Following [6], we call a semiring $S$ *atomistic* if whenever $x_1 + \cdots + x_m = y_1 + \cdots + y_n$ for some $x_1, \ldots, x_m, y_1, \ldots, y_n \in S$, where $m, n \geq 1$, then there exist $z_1, \ldots, z_k \in S$ and partitions $I_1, \ldots, I_m$ and $J_1, \ldots, J_n$ of the set $\{1, \ldots, k\}$ into pairwise disjoint (possibly empty) sets such that

$$x_i = \sum_{\ell \in I_i} z_\ell, \quad y_j = \sum_{\ell \in J_j} z_\ell$$

for all $1 \leq i \leq m$ and $1 \leq j \leq n$.

**Proposition 6.5** *Every atomistic semiring is equisubstractive.*

*Proof.* Suppose that $S$ is atomistic and $x + y = z + u$ for some $x, y, z, u \in S$. Then there exist $v_1, \ldots, v_k \in S$ and partitions $I_1, I_2$ and $J_1, J_2$ of the set $\{1, \ldots, k\}$ with $x = \sum_{\ell \in I_1} v_\ell$, $y = \sum_{\ell \in I_2} v_\ell$, $z = \sum_{\ell \in J_1} v_\ell$, $u = \sum_{\ell \in J_2} v_\ell$. Then let $a = \sum_{\ell \in I_1 \cap J_1} v_\ell$, $b = \sum_{\ell \in I_1 \cap J_2} v_\ell$, $c = \sum_{\ell \in I_2 \cap J_1} v_\ell$, $d = \sum_{\ell \in I_2 \cap J_2} v_\ell$. We clearly have that $x = a + b$, $y = c + d$, $z = a + c$, $u = b + d$. □

In [5], it is shown that every bounded distributive lattice is atomistic. It is clear that $\mathbb{N}$ is atomistic.

**Proposition 6.6** *Every ring is atomistic.*

*Proof.* Suppose that $R$ is a ring and $x_1 + \cdots + x_m = y_1 + \cdots + y_n$ for some $x_1, \ldots, x_m \in R$ and $y_1, \ldots, y_n \in R$. If $m = 1$ or $n = 1$, then it is clear that there exist $z_1, \ldots, z_k \in S$ and partitions $I_1, \ldots, I_m$ and $J_1, \ldots, J_n$ of the set $\{1, \ldots, k\}$ with the required properties. Suppose now that $m, n \geq 2$ and that our claim holds for all integers $m', n'$ with $m' + n' < m + n$. Then consider the sums

$$x_1 + \cdots + x_{m-1} + (x_m - y_n) = y_1 + \cdots + y_{n-1}.$$

By the induction hypothesis, we can find $z_1, \ldots, z_k \in R$ and partitions $I_1, \ldots, I_m$ and $J_1, \ldots, J_{n-1}$ of $\{1, \ldots, k\}$ with

$$x_i = \sum_{\ell \in I_i} z_\ell, \quad 1 \leq i < m$$

$$x_m - y_n = \sum_{\ell \in I_m} z_\ell$$

$$y_j = \sum_{\ell \in J_j} z_\ell, \quad 1 \leq j < n.$$

Then the family $z'_1 = z_1, \ldots, z'_k = z_k, z'_{k+1} = y_n$ and the partitions $I'_1 = I_1, \ldots, I'_{m-1} = I_{m-1}, I'_m = I_m \cup \{k+1\}, J'_1 = J_1, \ldots, J'_{n-1} = J_{n-1}, J'_n = \{k+1\}$ of the set $\{1, \ldots, k+1\}$ will do. □

## 7 Free iterative semialgebras

In this section, our principal aim is to prove the following result:

**Theorem 7.1** *Suppose that $K$ is a proper commutative semiring. Then for any finite alphabet $\Sigma$, $K^{\mathrm{rat}}\langle\!\langle \Sigma^* \rangle\!\rangle$ is a free iterative $K$-semialgebra on $\Sigma$. In more detail, if $A$ is any iterative $K$-semialgebra with distinguished $K$-ideal $I$, and if $h$ is any function $\Sigma \to I$, then there is a unique morphism of iterative $K$-semialgebras $h^\sharp : K^{\mathrm{rat}}\langle\!\langle \Sigma^* \rangle\!\rangle \to A$ extending $h$.*

*Proof.* Suppose that $A$ is an iterative $K$-semialgebra. Thus, $A$ has a distinguished ideal $I$ which is closed under the $K$-action such that for any $a \in I$ and $b \in A$, the equation $x = ax + b$ has a unique solution in $A$. Let $h$ be any function $h : \Sigma \to I$. We want to show that $h$ has a unique extension to a $K$-semialgebra morphism $K^{\mathrm{rat}}\langle\!\langle \Sigma^* \rangle\!\rangle \to A$. Such a morphism automatically preserves star.

To define $h^\sharp$, let us first extend $h$ to a linear mapping $K\Sigma \to A$ in the obvious way. If $r \in K^{\text{rat}}\langle\!\langle\Sigma^*\rangle\!\rangle$, there exists an automaton $\mathcal{A} = (\alpha, M, \beta)$ in $K\langle\!\langle\Sigma^*\rangle\!\rangle$ with $|\mathcal{A}| = r$. Let $\mathcal{A}h = (\alpha, Mh, \beta)$, where $Mh$ is defined pointwise. Then $\mathcal{A}h$ is an automaton in $A$ over $\Sigma h$. Finally, we define $rh^\sharp := |Mh|$. Using automata constructions, it can be shown that $h^\sharp$ preserves the $K$-semialgebra operations, see [6], Theorem 5.1 for details. Thus, if $h^\sharp$ is well-defined, then $h^\sharp$ is a morphism of iterative $K$-semialgebras extending $h$. So it remains to show that $h^\sharp$ is a function. But since $K$ is proper, it suffices to show that if $X$ is a simulation $\mathcal{A} \to \mathcal{B}$, then $\mathcal{A}h$ and $\mathcal{B}h$ are equivalent.

So suppose that $X \in K^{m \times n}$ is a simulation $\mathcal{A} \to \mathcal{B}$, where $\mathcal{A} = (\alpha, M, \beta)$ and $\mathcal{B} = (\gamma, N, \delta)$ with $M \in (K\Sigma)^{m \times m}$ and $N \in (K\Sigma)^{n \times n}$. Then $X$ is a simulation $\mathcal{A}h \to \mathcal{B}h$, where $\mathcal{A}h = (\alpha, Mh, \beta)$ and $\mathcal{B} = (\gamma, Nh, \delta)$, as the reader can easily verify. Now since $A$ is an iterative $K$-semialgebra, by Lemma 5.2 we have that $\mathcal{A}h$ and $\mathcal{B}h$ are equivalent. $\square$

**Corollary 7.2** *Suppose that $K$ is a proper commutative semiring. Then for any finite alphabet $\Sigma$, $K^{\text{rat}}\langle\!\langle\Sigma^*\rangle\!\rangle$ is a free symmetric iterative $K$-semialgebra on $\Sigma$.*

*Proof.* This follows from Theorem 7.1 by noting that $K^{\text{rat}}\langle\!\langle\Sigma^*\rangle\!\rangle$ is a symmetric iterative semiring. $\square$

**Corollary 7.3** *Suppose that $K$ is a commutative semiring such that each finitely generated subsemiring of $K$ is contained in a Noetherian subsemiring. Then for any finite alphabet $\Sigma$, $K^{\text{rat}}\langle\!\langle\Sigma^*\rangle\!\rangle$ is a free (symmetric) iterative $K$-semialgebra on $\Sigma$.*

Theorem 7.1 and Corollary 7.2 apply to all commutative rings, all finite commutative semirings, and the semiring $\mathbb{N}$. Corollary 7.3 applies to all commutative rings and all finite commutative semirings.

**Corollary 7.4** *Suppose that $K$ is a commutative ring. Then for any finite alphabet $\Sigma$, $K^{\text{rat}}\langle\!\langle\Sigma^*\rangle\!\rangle$ has the following universal property: Given any $K$-semialgebra $A$ with a distinguished $K$-ideal $I$ such that for each $a \in I$ the element $e - a$ has an inverse $a^* \in A$, and given any function $h : \Sigma \to I$, there is a unique $K$-semialgebra morphism $K^{\text{rat}}\langle\!\langle\Sigma^*\rangle\!\rangle \to A$ which extends $h$ (and preserves $^*$).*

*Proof.* By Corollary 7.3 and Example 3.4. $\square$

Corollary 7.4 is closely related to result in [22, 3].

## 8 Free partial iteration semialgebras

In order to give a complete equational axiomatization of the algebra of regular languages, Conway associated an identity with each finite group, cf. [8].

Suppose that $G$ is a finite group of order $n$. Without loss of generality we may assume that the elements of $G$ are the integers in $[n] = \{1, \ldots, n\}$ with corresponding variables $x_1, \ldots, x_n$ that will be interpreted as elements of the distinguished ideal of a partial Conway semiring. Let · denote the product operation of $G$. The structure of $G$ can be fully described by an $n \times n$ matrix $M_G$ whose $(i,j)$th entry is the variable $x_{i^{-1} \cdot j}$. Each row and each column of $M_G$ is a permutation of the first row. Let us define $M_G^*$ by the matrix star formula (4). Note that each entry of $M_G^*$ is a term in the variables $x_1, \ldots, x_n$, composed by the semiring operations and star.

Let $r_1, \ldots, r_n$ denote the terms appearing in the first row of $M_G^*$. It is known (cf. [8, 23]) that, modulo the identities (2) and (3), each row and each column of $M_G^*$ is a permutation of $r_1, \ldots, r_n$. The *group identity associated with the finite group $G$* is

$$r_1 + \cdots + r_n \ = \ (x_1 + \cdots + x_n)^*. \tag{5}$$

Using matrix notation, the group identity associated with $G$ can be written as

$$e_1 M_G^* u_n \ = \ (x_1 + \cdots + x_n)^*,$$

where $e_1$ is a $n$-dimensional row vector whose first component is 1 and whose other components are 0, and $u_n$ is an $n$-dimensional column vector whose components are all 1.

In [6], a *partial iteration semiring* is defined to be a partial Conway semiring $S$ with distinguished ideal $I$ that satisfies all group identities (5) when the variables $x_1, \ldots, x_n$ are interpreted as elements of $I$. Morphisms of partial iteration semirings are partial Conway semiring morphisms.

**Definition 8.1** *Let $K$ be a commutative semiring and $(A, I, ^*)$ a partial Conway $K$-semialgebra. We call $A$ a* partial iteration $K$-semialgebra *if $A$, equipped with the $K$-ideal $I$, is a partial iteration semiring. A morphism of partial iteration $K$-semialgebras is a partial Conway $K$-semialgebra morphism.*

For the following result, see [6].

**Theorem 8.2** *Any iterative $K$-semialgebra is a partial iteration $K$-semialgebra. Any morphism of iterative $K$-semialgebras is a partial iteration $K$-semialgebra morphism.*

Thus, when $K$ is a commutative semiring and $\Sigma$ is a finite alphabet, then $K\langle\!\langle \Sigma^* \rangle\!\rangle$ with distinguished ideal the collection of all proper series, is a partial iteration $K$-semialgebra. Moreover, $K^{\mathrm{rat}}\langle\!\langle \Sigma^* \rangle\!\rangle$, with distinguished ideal the collection of all proper rational series, is also a partial iteration $K$-semialgebra.

In this section, our aim is to describe the structure of free partial iteration $K$-semialgebras for certain commutative semirings $K$. Before presenting this result, we need some preparations.

**Lemma 8.3** *Let $K$ be a commutative semiring and $(A, I, ^*)$ a partial Conway $K$-semialgebra. Suppose that $M \in I^{n \times n}$ and that $X \in K^{n \times n}$ is an invertible diagonal matrix. Then $(X^{-1}MX)^* = X^{-1}M^*X$.*

*Proof.* This is clear when $n = 1$. Suppose now that $n > 1$ and write

$$M = \begin{pmatrix} a & b \\ c & d \end{pmatrix} \quad X = \begin{pmatrix} k & 0 \\ 0 & \ell \end{pmatrix}$$

where $a$ is an $(n-1) \times (n-1)$, $d$ is a $1 \times 1$ matrix, etc. Our assumption is that $k$ and $\ell$ are invertible diagonal matrices with inverses $k^{-1}$ and $\ell^{-1}$. Now by the matrix star formula,

$$(X^{-1}MX)^* = \begin{pmatrix} k^{-1}ak & k^{-1}b\ell \\ \ell^{-1}ck & \ell^{-1}d\ell \end{pmatrix}^* = \begin{pmatrix} x & y \\ z & u \end{pmatrix}$$

where, using the induction hypothesis,

$$\begin{aligned} x &= (k^{-1}ak + k^{-1}b\ell(\ell^{-1}d\ell)^*\ell^{-1}ck)^* \\ &= (k^{-1}(a + bd^*c)k)^* \\ &= k^{-1}(a + bd^*c)^*k \\ &= k^{-1}x'k \\ y &= k^{-1}(a + bd^*c)^*kk^{-1}b\ell(\ell^{-1}d\ell)^* \\ &= k^{-1}(a + bd^*c)^*bd^*\ell \\ &= k^{-1}y'\ell, \end{aligned}$$

and similarly,

$$\begin{aligned} u &= \ell^{-1}(d + ca^*b)^*\ell = \ell^{-1}u'\ell \\ z &= \ell^{-1}(d + ca^*b)^*ca^*k = \ell^{-1}z'k. \end{aligned}$$

Thus,

$$(X^{-1}MX)^* = \begin{pmatrix} k^{-1}x'k & k^{-1}y'\ell \\ \ell^{-1}z'k & \ell^{-1}u'\ell \end{pmatrix} = X^{-1}M^*X. \quad \square$$

**Lemma 8.4** *Let $K$ be a strongly proper atomistic commutative semiring and $(A, I, ^*)$ a partial iteration $K$-semialgebra. Let $M \in I^{m \times m}$, $N \in I^{n \times n}$. Suppose that $X \in K^{m \times n}$ is a functional or dual functional matrix with $MX = XN$. Then $M^*X = XN^*$.*

This follows from [5], Corollaries 19, 20 and Proposition 15. We are now ready to prove the main result of this section.

**Theorem 8.5** *Suppose that $K$ is a strongly proper atomistic commutative semiring. Then for each finite alphabet $\Sigma$, $K^{\mathrm{rat}}\langle\!\langle \Sigma^* \rangle\!\rangle$ is a free partial iteration $K$-semialgebra on $\Sigma$.*

*Proof.* Suppose that $(A, I, ^*)$ is a partial iteration $K$-semialgebra and let $h$ be a function $\Sigma \to A$, where $\Sigma$ is a finite alphabet. We extend $h$ to a morphism $h^\sharp : K^{\mathrm{rat}}\langle\!\langle \Sigma^* \rangle\!\rangle \to A$ as in the proof of Theorem 7.1: For all $s = |\mathcal{A}|$, where $\mathcal{A}$ is an automaton in $K\langle\!\langle \Sigma^* \rangle\!\rangle$, we define $sh^\sharp = |\mathcal{A}h|$. Since $K$ is strongly proper, it follows that $h^\sharp$ is well-defined if we can show that whenever there is a functional, dual functional or invertible diagonal simulation $\mathcal{A} \to \mathcal{B}$, then $|\mathcal{A}h| = |\mathcal{B}h|$.

Let $\mathcal{A} = (\alpha, M, \beta)$, $\mathcal{B} = (\gamma, N, \delta)$ be automata in $K\langle\!\langle \Sigma^* \rangle\!\rangle$ of dimension $m$ and $n$, and let $X \in K^{m \times n}$. Suppose first that $X \in K^{m \times n}$ is a diagonal invertible simulation $\mathcal{A} \to \mathcal{B}$, so that $m = n$. Then, using Lemma 8.3,

$$|\mathcal{A}h| = \alpha(Mh)^*\beta = \alpha(Mh)^*X\delta = \alpha X(Nh)^*\delta = \gamma(Nh)^*\delta = |\mathcal{B}h|.$$

When $X$ is a functional simulation or a dual functional simulation, then the proof of $|\mathcal{A}h| = |\mathcal{B}h|$ is the same, using Lemma 8.4. $\square$

**Corollary 8.6** *Suppose that $K$ is a proper atomistic commutative semiring which is additively generated by its units. Then for each finite alphabet $\Sigma$, $K^{\mathrm{rat}}\langle\!\langle \Sigma^*\rangle\!\rangle$ is a free partial iteration $K$-semialgebra on $\Sigma$.*

The above corollary applies to all commutative rings, the semiring $\mathbb{N}$, and the semirings $\mathbb{N}_k$, $k \geq 2$ defined above, obtained from $\mathbb{N}$ by collapsing all integers greater than or equal to $k-1$ to a single element.

**Corollary 8.7** *Suppose that $K$ is a strongly proper commutative semiring. Then for each finite alphabet $\Sigma$, $K^{\mathrm{rat}}\langle\!\langle \Sigma^*\rangle\!\rangle$ is a free partial iteration $K$-semialgebra on $\Sigma$ in the class of all partial Conway $K$-semialgebras $(A, I,^*)$ satisfying the following two conditions: For all $X \in I^{m \times m}$ and $Y \in I^{1 \times 1}$, if $X\rho = \rho Y$ then $X^*\rho = \rho Y^*$, and if $Y\rho^T = \rho^T X$ then $Y^*\rho^T = \rho^T X^*$, where $\rho$ denotes the unique functional matrix in $K^{m \times 1}$ and $\rho^T$ is its transpose.*

*Proof.* On the one hand, $K^{\mathrm{rat}}\langle\!\langle \Sigma^*\rangle\!\rangle$ is an iterative $K$-semialgebra and thus satisfies the above implications.

It is known that if the implications

$$X\rho = \rho Y \Longrightarrow X^*\rho = \rho Y^*$$

and

$$Y\rho^T = \rho^T X \Longrightarrow Y^*\rho^T = \rho^T X^*$$

hold for the unique functional matrix $\rho \in K^{m \times 1}$ and for all $X \in I^{m \times m}$ and $Y \in I^{1 \times 1}$, where $(A, I,^*)$ is a partial Conway $K$-semialgebra, then the same implications hold for all $X \in I^{m \times m}$, $Y \in I^{n \times n}$, and for all functional matrices $\rho$ in $K^{m \times n}$. See [5]. □

## 9 Iteration $K$-semialgebras

In this section, we consider iteration $K$-semialgebras with a completely defined star operation. We will assume that the commutative semiring $K$ is also equipped with a star operation, and that the two star operations are compatible.

We recall from [4] that a *Conway semiring* is a partial Conway semiring $S$ with a totally defined star operation. Thus, the distinguished ideal is $S$. Similarly, an *iteration semiring* is a partial iteration semiring with a totally defined star operation. Morphisms of Conway and iteration semirings preserve star.

**Definition 9.1** *Suppose that $K$ is a commutative semiring equipped with a star operation $^*: K \to K$. We call a $K$-semialgebra $A$ a* Conway $K$-semialgebra *if $A$ is a Conway semiring and the identity*

$$(ke)^* = k^*e \tag{6}$$

*holds for all $k \in K$. If additionally $A$ satisfies the group identities, then $A$ is an* iteration $K$-semialgebra. *A morphism of Conway or iteration $K$-semialgebras is a $K$-semialgebra morphism that preserves star.*

**Remark 9.2** *When $K$ and $A$ are Conway semirings, (6) is equivalent to the equation*

$$(ke)^+ = k^+e, \tag{7}$$

*for all $k \in K$. Indeed, if (6) holds, then $(ke)^+ = (ke)(ke)^* = (ke)(k^*e) = k^+e$. And if (7) holds, then $(ke)^* = (ke)^+ + e = k^+e + e = k^*e$.*

**Remark 9.3** *If $A$ is an iteration $K$-semialgebra, where $K$ is a commutative semiring equipped with a star operation, and if the function $K \to A$, $k \mapsto ke$ is injective, as in the case when $A = K\langle\!\langle \Sigma^* \rangle\!\rangle$ or $A = K^{\mathrm{rat}}\langle\!\langle \Sigma^* \rangle\!\rangle$ for some $\Sigma$, then $K$ is also an iteration semiring.*

The following fact was proved in [4].

**Theorem 9.4** *If $S$ is a (not necessarily commutative) iteration semiring, then for each finite alphabet $\Sigma$, $S\langle\!\langle \Sigma^* \rangle\!\rangle$ is an iteration semiring in a unique way such that the star operation extends to one defined on $S$.*

In particular, if $K$ is a commutative iteration semiring, then $K\langle\!\langle \Sigma^* \rangle\!\rangle$ is an iteration $K$-semialgebra. On proper series, the star operation agrees with the one defined in Section 3. In the same way, $K^{\mathrm{rat}}\langle\!\langle \Sigma^* \rangle\!\rangle$ is an iteration semiring, since if $s$ is a rational series, then so is $s^*$. Indeed, if $s \in K^{\mathrm{rat}}\langle\!\langle \Sigma^* \rangle\!\rangle$, then $s = k_0\epsilon + r$, where $k_0 \in K$ and $r$ is a proper rational series. Now by (2), $s^* = (k_0^*r)^*k_0^* = k_0^*\epsilon + (k_0^*r)^+k_0^*$, showing that $s^*$ is also rational.

**Theorem 9.5** *Suppose that $K$ is a strongly proper and atomistic commutative iteration semiring. Then for each finite alphabet $\Sigma$, $K^{\mathrm{rat}}\langle\!\langle \Sigma^* \rangle\!\rangle$ is a free iteration $K$-semialgebra on $\Sigma$.*

*Proof.* Let $A$ be an iteration $K$-semialgebra and $h$ a function $\Sigma \to A$. Since any iteration $K$-semialgebra is a partial iteration $K$-semialgebra, from Theorem 8.5 we know that $h$ can be extended to a unique $K$-semialgebra morphism $h^\sharp : K^{\mathrm{rat}}\langle\!\langle \Sigma^* \rangle\!\rangle \to A$ that preserves the star operation on proper rational series. So it remains to show that $h^\sharp$ preserves the star operation on non-proper series. But any series $s \in K^{\mathrm{rat}}\langle\!\langle \Sigma^* \rangle\!\rangle$ can be written as $k_0\epsilon + r$, where $k_0 \in K$ and $r$ is a proper rational series. By (2), we have $s^* = (k_0^*r)^*k_0^*$. Since $k_0^*r$ is proper and rational, $(k_0^*r)^*h^\sharp = (k_0^*(rh^\sharp))^*$. Since $h^\sharp$ preserves the action, also $((k_0^*r)^*k_0^*)h^\sharp = (k_0^*(rh^\sharp))^*k_0^*$. Thus, $s^*h^\sharp = ((k_0^*r)^*k_0^*)h^\sharp = (k_0^*(rh^\sharp))^*k_0^* = ((k_0^*e)(rh^\sharp))^*(k_0^*e) = ((k_0e)^*(rh^\sharp))^*(k_0e)^* = (k_0e + rh^\sharp)^* = ((k_0\epsilon + r)h^\sharp)^* = (sh^\sharp)^*$, proving that $h^\sharp$ preserves star. $\square$

The above result applies to the semirings $\mathbb{N}_k$, $k \geq 2$, with star operation $0^* = 1$ and $n^* = k-1$, for all $0 < n < k$. The case $k = 2$ was considered in [23].

## 10 An adjunction

When $S$ is a semiring and $\infty \notin S$, then we extend the sum and product operations to the set $S_\infty = S \cup \{\infty\}$ by defining $\infty + x = x + \infty = \infty$ and $y\infty = \infty y = \infty$ for all $x, y \in S_\infty$, $y \neq 0$.

It is known, cf. [20], that $S_\infty$ is itself a semiring iff $S$ is both *zerosum-free* and *zerodivisor-free*, i.e., when

$$x + y = 0 \implies x = y = 0$$
$$xy = 0 \implies x = 0 \text{ or } y = 0$$

for all $x, y \in S$. We call a semiring $S$ *positive* if it is both zerosum-free and zerodivisor-free. Note that $S$ is positive iff the function $S \to \mathbb{B}$, $0 \mapsto 0$, $x \mapsto 1$ for all $x \neq 0$ is a semiring morphism.

**Proposition 10.1** *Suppose that $S$ is a Conway semiring. Then $S$ is zerosum-free.*

*Proof.* Let $x := 1^*$. Then $x + 1 = x$. By Proposition 1.23 in [19], it follows that $S$ is zerosum-free. $\square$

**Proposition 10.2** *Let $S$ be a positive semiring. Then $S$ can be embedded in an iteration semiring.*

*Proof.* Consider the semiring $S_\infty$ that contains $S$ as a subsemiring. For each family $(s_i)_{i \in I}$ of elements of $S_\infty$, where $I$ is any index set, define $\sum_{i \in I} s_i := s_{i_1} + \ldots + s_{i_k}$ if $i_1, \ldots, i_k$ are all the elements $j \in I$ with $s_j \neq 0$, and define $\sum_{i \in I} s_i := \infty$ otherwise. Equipped with this summation, $S_\infty$ is a *complete semiring* cf. [11, 18, 20]. Now define $x^* = \sum_{n \geq 0} x^n$ for all $x \in S_\infty$. It is known that equipped with this star operation, every complete semiring is an iteration semiring. Note that $0^* = 1$ and $x^* = \infty$ for all $x \neq 0$. $\square$

In the rest of this section, we assume that $K$ is a positive commutative semiring so that $K_\infty$ is turned into an iteration semiring. We will also assume a basic knowledge of variety theory as presented e.g. in [7].

Let $\mathcal{V}$ be a subvariety of Conway $K_\infty$-semialgebras satisfying

$$(\infty a)^+ = \infty a^+, \tag{8}$$

for all $a \in A$. Moreover, let $\mathcal{W}$ be the subvariety of $\mathcal{V}$ formed by those Conway $K_\infty$-semialgebras $A \in \mathcal{V}$ satisfying

$$\infty e = e. \tag{9}$$

It follows that for all $k \in K_\infty$, $k \neq 0$,

$$ke = k(\infty e) = (k\infty)e = \infty e = e$$

holds in $\mathcal{W}$. Moreover, if $A \in \mathcal{W}$, then

$$ka = k(ea) = (ke)a = ea = a$$

holds for all $a \in A$ and $k \in K_\infty$, $k \neq 0$. In particular, $a + a = (1 + 1)a = a$, for all $a \in A$, showing that any $A \in \mathcal{W}$ is idempotent.

**Remark 10.3** *Suppose that $A \in \mathcal{W}$. Since $K_\infty$ is a Conway semiring, (7) holds. Thus, $e^+ = \infty e^+ = (\infty e)^+ = \infty^+ e = e$. Also, $e^* = e^+ + e = e + e = e$.*

For later use we note:

**Lemma 10.4** *The identity $\infty(\infty a)^* = \infty a^*$ holds in $\mathcal{V}$.*

*Proof.* $\infty(\infty a)^* = \infty e + \infty(\infty a)^+ = \infty e + \infty a^+ = \infty a^*$. □

Both $\mathcal{V}$ and $\mathcal{W}$ give rise to categories whose morphisms are Conway $K$-semialgebra morphisms. It is well-known that the inclusion functor $\mathcal{W} \hookrightarrow \mathcal{V}$ has a left adjoint. In this section our aim is to provide a concrete description of this left adjoint.

Let $A \in \mathcal{V}$. Then we define $A\kappa := \{\infty a : a \in A\}$. The set $A\kappa$ contains 0 and is closed under sum, product and the $K_\infty$-action. Also, $\infty e \in A\kappa$, and $(\infty e)(\infty a) = (\infty\infty)(ea) = \infty a$ for all $a \in A$. Similarly, $(\infty a)(\infty e) = \infty a$. Thus, $(A\kappa, +, \cdot, 0, \infty e)$, equipped with the $K_\infty$-action inherited from $A$ is a $K_\infty$-semialgebra.

Also, for any $a \in A$, $(\infty a)^+ = \infty a^+ \in A\kappa$. We define a star operation on $A\kappa$ by $(\infty a)^\otimes := \infty e + (\infty a)^+ = \infty e + \infty a^+ = \infty a^*$, for all $a \in A$. This definition is valid, since if $\infty a = \infty b$, then $(\infty a)^\otimes = \infty e + (\infty a)^+ = \infty e + (\infty b)^+ = (\infty b)^\otimes$.

**Proposition 10.5** *Equipped with the above operations, $A\kappa$ is a Conway $K_\infty$-semialgebra in $\mathcal{W}$ that is a quotient of $A$. A surjective $K_\infty$-semialgebra morphism $A \to A\kappa$ is the map $\kappa : a \mapsto \infty a$, for all $a \in A$.*

*Proof.* Clearly, $A\kappa \in \mathcal{W}$. It is a routine matter to show that $\kappa$ preserves all operations and constants. We only show that $\kappa$ preserves star.

Let $a \in A$. Then, by the above definition, $(a^*\kappa) = \infty a^* = (\infty a)^\otimes = (a\kappa)^\otimes$. □

**Proposition 10.6** *Suppose that $A \in \mathcal{V}$, $A' \in \mathcal{W}$ and $h : A \to A'$ is a Conway $K$-semialgebra morphism. Then $h$ factors through $\kappa : A \to A\kappa$.*

*Proof.* Suppose that $\infty a = a\kappa = b\kappa = \infty b$ holds for some $a, b \in A$. Then $ah = \infty(ah) = (\infty a)h = (\infty b)h = \infty(bh) = (\infty b)h$, showing that the kernel of $\kappa$ is included in the kernel of $h$. □

We summarize the results of this section:

**Theorem 10.7** *The functor mapping $A \in \mathcal{V}$ to $A\kappa \in \mathcal{W}$ and a Conway $K_\infty$-semialgebra morphism $h : A \to B$ with $A, B \in \mathcal{V}$ to its restriction to $A\kappa$ is a left adjoint of the inclusion functor $\mathcal{W} \hookrightarrow \mathcal{V}$.*

A *Conway $K_\infty$-semialgebra term* $t = t(x_1, \ldots, x_n)$ in the variables $x_1, \ldots, x_n$ is defined as expected: $x_1, \ldots, x_n$ are terms as are 0 and $e$, and if $t, t_1, t_2$ are terms and $k \in K_\infty$, then $t_1 + t_2$, $t_1 t_2$, $kt$ and $t^*$ are also terms. We write $t^+$ as an abbreviation for $tt^*$. An identity (or equation) is a formal equality between two terms. The definition of whether an identity holds in a Conway $K_\infty$-semialgebra is standard.

**Corollary 10.8** *Suppose that $s = s(x_1, \ldots, x_n)$ and $t = t(x_1, \ldots, x_n)$ are Conway $K_\infty$-semi-algebra terms in the variables $x_1, \ldots, x_n$. Then $\infty s = \infty t$ holds in $\mathcal{V}$ iff it holds in $\mathcal{W}$.*

*Proof.* If $\infty s = \infty t$ holds in $\mathcal{V}$ then it clearly holds in $\mathcal{W}$, since $\mathcal{W} \subseteq \mathcal{V}$. Suppose now that $\infty s = \infty t$ holds in $\mathcal{W}$. Then, for any $A \in \mathcal{V}$ and $a_1, \ldots, a_n \in A$,

$$\begin{aligned}
(\infty s)^A(a_1, \ldots, a_n) &= \infty s^A(a_1, \ldots, a_n) \\
&= s^A(a_1, \ldots, a_n)\kappa \\
&= s^{A\kappa}(a_1\kappa, \ldots, a_n\kappa) \\
&= t^{A\kappa}(a_1\kappa, \ldots, a_n\kappa) \\
&= t^A(a_1, \ldots, a_n)\kappa \\
&= \infty t^A(a_1, \ldots, a_n) \\
&= (\infty t)^A(a_1, \ldots, a_n).
\end{aligned}$$

$\square$

## 11 Iteration semialgebras, revisited

In this section, we fix a commutative positive semiring $K$ and consider the iteration semiring $K_\infty$. Note that $K_\infty$ is also commutative and positive. The main result of this section is:

**Theorem 11.1** *For each finite alphabet $\Sigma = \{\sigma_1, \ldots, \sigma_n\}$, $K_\infty^{\mathrm{rat}}\langle\!\langle \Sigma^* \rangle\!\rangle$ is freely generated by $\Sigma$ in the variety $\mathcal{V}$ of all iteration $K_\infty$-semialgebras satisfying (8).*

It is clear that $K_\infty^{\mathrm{rat}}\langle\!\langle \Sigma^* \rangle\!\rangle$ satisfies (8) and is generated by $\Sigma$. Thus, it suffices to show that whenever $s = s(x_1, \ldots, x_n)$ and $t = t(x_1, \ldots, x_n)$ are iteration $K_\infty$-semialgebra terms in the variables $x_1, \ldots, x_n$ with $|t| = |s|$, then $t = s$ holds in $\mathcal{V}$, where $|t| = t^{K_\infty^{\mathrm{rat}}\langle\!\langle \Sigma^* \rangle\!\rangle}(\sigma_1, \ldots, \sigma_n)$ and $|s|$ is defined in the same way.

Since $K$ is positive, the support of any series in $K_\infty^{\mathrm{rat}}\langle\!\langle \Sigma^* \rangle\!\rangle$ is regular and the map that takes a series $s \in K_\infty^{\mathrm{rat}}\langle\!\langle \Sigma^* \rangle\!\rangle$ to the characteristic series of $\mathrm{supp}(s)$ is an iteration semiring morphism $K_\infty^{\mathrm{rat}}\langle\!\langle \Sigma^* \rangle\!\rangle \to \mathbb{B}^{\mathrm{rat}}\langle\!\langle \Sigma^* \rangle\!\rangle$. We can turn $\mathbb{B}^{\mathrm{rat}}\langle\!\langle \Sigma^* \rangle\!\rangle$ into an iteration $K_\infty$-semialgebra by defining $0r := 0$ and $kr := r$, for all $r \in \mathbb{B}^{\mathrm{rat}}\langle\!\langle \Sigma^* \rangle\!\rangle$ and $k \in K_\infty$, $k \neq 0$. Note that $\mathbb{B}^{\mathrm{rat}}\langle\!\langle \Sigma^* \rangle\!\rangle$ is in the subvariety $\mathcal{W}$ as defined in Section 10. Moreover, the above map is an iteration $K_\infty$-semialgebra morphism.

**Lemma 11.2** *For each finite alphabet $\Sigma$, $\mathbb{B}^{\mathrm{rat}}\langle\!\langle \Sigma^* \rangle\!\rangle$ is freely generated by $\Sigma$ in $\mathcal{W}$.*

*Proof.* Suppose that $A$ is an iteration $K_\infty$-semialgebra in $\mathcal{W}$ and $h : \Sigma \to A$ is a mapping. Then $A$ is naturally an iteration $\mathbb{B}$-semialgebra and by Theorem 9.5, $h$ extends to a unique morphism of iteration $\mathbb{B}$-semialgebras $\mathbb{B}^{\mathrm{rat}}\langle\!\langle \Sigma^* \rangle\!\rangle \to A$. To complete the proof, it suffices to show that when $\mathbb{B}^{\mathrm{rat}}\langle\!\langle \Sigma^* \rangle\!\rangle$ and $A$ are viewed as $K_\infty$- semialgebras, then $h^\sharp$ preserves the $K_\infty$-action. To this end, let $r \in \mathbb{B}^{\mathrm{rat}}\langle\!\langle \Sigma^* \rangle\!\rangle$ and $k \in K_\infty$. If $k \neq 0$, then $h^\sharp(kr) = h^\sharp(r) = kh^\sharp(r)$. If $k = 0$, then $h^\sharp(kr) = 0 = kh^\sharp(r)$. $\square$

Below we will say that terms $s, t$ are *equivalent* if $s = t$ holds in $\mathcal{V}$. We will make use of several lemmas. A constant term is a term with no variables.

The following fact is clear:

**Lemma 11.3** *For every term $t$ there is a constant term $t_c$ and a term $t'$ such that $t$ and $t_c + t'$ are equivalent and $|t'|$ is proper.*

**Lemma 11.4** *Suppose that $t$ is a constant term. Then there is some $k \in K_\infty$ such that $t$ is equivalent to the term $ke$.*

*Proof.* This follows by noting that $0$ is equivalent to $0e$, $e$ is equivalent to $1e$, and that $ke + k'e$ is equivalent to $(k + k')e$, $(ke) \cdot (k'e)$ is equivalent to $(kk')e$, $k(k'e)$ is equivalent to $(kk')e$, moreover, $(ke)^*$ is equivalent to $k^*e$. □

By this lemma, we may assume that each constant term is either $0$ or a term of the form $ke$, where $k \neq 0$. Below, to increase readability, we will identify terms of the form $(ke)t$ and $kt$ and denote terms $t(ke)$ by just $tk$ (recall that the left action induces a right action), and moreover, we will sometimes write just $k$ for $ke$.

Call a term $t$ *simple* if it can be constructed from the variables and the term $0$ by the operations of sum, product, $K$-action and plus, where $t^+$ is an abbreviation for $tt^*$. Note that if $t$ is a simple term in the variables $x_1, \ldots, x_n$, then $|t|$ is a proper series all of whose coefficients are in $K$.

The following lemma is an extension of [5], Lemma 39, that covers the case $K = \mathbb{N}$.

**Lemma 11.5** *Each term $t$ in the variables $x_1, \ldots, x_n$ is equivalent to a term of the form $t_c + t_0 + \infty t_\infty$, where $t_c$ is a constant term and $t_0$ is simple.*

*Proof.* This is clear when the term is $0$, $e$, or a variable. Supposing that the claim holds for terms $s$ and $t$, we show it holds for the terms $s + t$, $st$ and $s^*$ and $ks$, where $k \in K_\infty$.

*Case of $s + t$.* Clearly, $s + t$ is equivalent to $(s_c + t_c) + (s_0 + t_0) + \infty(s_\infty + t_\infty)$.

*Case of $st$.* If $s_c, t_c \in K$ then $st$ is equivalent to $s_c t_c + (s_c t_0 + s_0 t_c + s_0 t_0) + \infty((s_c + s_0)t_\infty + s_\infty(t_c + t_0) + s_\infty t_\infty)$. If $s_c = \infty$ but $t_c \in K$, then $st$ is equivalent to $s_c t_c + (s_0 t_c + s_0 t_0) + \infty(t_0 + (s_c + s_0)t_\infty + s_\infty(t_c + t_0) + s_\infty t_\infty)$. The case when $s_c \in K$ and $t_c = \infty$ is symmetric. Finally, when $s_c = t_c = \infty$ then $st$ is equivalent to $\infty + s_0 t_0 + \infty(s_0 + t_0 + (s_c + s_0)t_\infty + s_\infty(t_c + t_0) + s_\infty t_\infty)$.

*Case of $s^*$.* If $s_c = 0$, then $s^*$ is equivalent to $e + s_0^+ + \infty(s_0 + s_\infty)^* s_\infty s_0^*$ as shown by the following computation using the sum star and product star identities and Lemma 10.4.

$$\begin{aligned}
(s_0 + \infty s_\infty)^* &= s_0^*(\infty s_\infty s_0^*)^* \\
&= s_0^*(1 + \infty(s_\infty s_0^* \infty)^* s_\infty s_0^*) \\
&= s_0^* + s_0^* \infty(\infty s_\infty s_0^*)^* s_\infty s_0^* \\
&= s_0^* + s_0^* \infty(s_\infty s_0^*)^* s_\infty s_0^* \\
&= s_0^* + \infty s_0^*(s_\infty s_0^*)^* s_\infty s_0^* \\
&= e + s_0^+ + \infty(s_0 + s_\infty)^* s_\infty s_0^*.
\end{aligned}$$

If $s_c \neq 0$, then $s_c^*$ is equivalent to $\infty e$ and thus $s^*$ is equivalent to

$$\infty((s_0 + s_\infty)\infty)^* = \infty(s_0 + s_\infty)^* = \infty e + \infty(s_0 + s_\infty)^+.$$

In either case, $s^*$ is of the required form.

Suppose now that the claim holds for the term $s$ and let $k \in K_\infty$. If $k = 0$ then $ks$ is equivalent to the term $0$ and our claim is clear. If $k \in K$, $k \neq 0$, then $ks$ is equivalent to $ks_c + ks_0 + \infty s_\infty$, since $k\infty = \infty$. Last, if $k = \infty$, then $ks$ is equivalent to $\infty(s_c + s_0 + s_\infty)$. □

Using Lemma 11.3, we have:

**Corollary 11.6** *Each term $t$ in the variables $x_1, \ldots, x_n$ is equivalent to a term of the form $t_c + t_0 + \infty t_\infty$, where $t_c$ is a constant term and $t_0$ is simple, moreover, either $t_c = 0$ or $|t_\infty|$ is proper.*

**Corollary 11.7** $K_\infty^{\mathrm{rat}}\langle\!\langle \Sigma^* \rangle\!\rangle$ *is a Fatou extension [3] of $K^{\mathrm{rat}}\langle\!\langle \Sigma^* \rangle\!\rangle$, i.e., $K_\infty^{\mathrm{rat}}\langle\!\langle \Sigma^* \rangle\!\rangle \cap K\langle\!\langle \Sigma^* \rangle\!\rangle = K^{\mathrm{rat}}\langle\!\langle \Sigma^* \rangle\!\rangle$.*

By different methods, we can show that this fact holds for positive semirings $K$ that are not necessarily commutative.

**Lemma 11.8** *Suppose that $s$ and $t$ are terms in the variables $x_1, \ldots, x_n$ such that $\mathrm{supp}(|s|) \subseteq \mathrm{supp}(|t|)$. Then $s + \infty t$ is equivalent to $\infty t$.*

*Proof.* Since $\mathrm{supp}(|\infty s|) = \mathrm{supp}(|s|)$, we also have $\mathrm{supp}(|\infty s|) \subseteq \mathrm{supp}(|t|)$ and $\mathrm{supp}(|\infty(s+t)|) = \mathrm{supp}(|\infty t|)$. This means that $\infty(s+t)$ and $\infty(t)$ evaluate to the same series in $\mathbb{B}^{\mathrm{rat}}\langle\!\langle \Sigma^* \rangle\!\rangle$ when each variable $x_i$ is interpreted as the letter $\sigma_i$. By Lemma 11.2, this implies that $\infty(s+t) = \infty t$ holds in $\mathcal{W}$ and thus by Corollary 10.8, $\infty(s+t)$ and $\infty t$ are equivalent. Thus,

$$s + \infty t = s + \infty s + \infty t = (1 + \infty)s + \infty t = \infty s + \infty t = \infty t$$

holds in $\mathcal{V}$. □

Next we give a generalization of [5], Lemma 43.

**Lemma 11.9** *For every term $t$ in the variables $x_1, \ldots, x_n$ there is an equivalent term of the form $t_c + t_0 + \infty t_\infty$ such that $t_c$ is a constant term, $t_0$ is simple, and the supports of $|t_c|$, $|t_0|$ and $|t_\infty|$ are pairwise disjoint.*

*Proof.* By Corollary 11.6, $t$ is equivalent to a term of the form $t_c + t_0 + \infty t_\infty$, where $t_0, t_c$ and $t_\infty$ satisfy the required conditions except possibly that the supports of $|t_0|$ and $|t_\infty|$ are not disjoint. Since $\mathrm{supp}(|t_\infty|)$ is a regular language, and since the "Hadamard product" of rational series with coefficients in a commutative semiring is rational, cf. [31] or [3], Theorem 5.5, we can write $|t_0|$ as a sum $s_1 + s_2$, where $s_1, s_2$ are proper series in $K^{\mathrm{rat}}\langle\!\langle \Sigma^* \rangle\!\rangle$ with $\mathrm{supp}(|s_1|) \cap \mathrm{supp}(|t_\infty|) = \emptyset$ and $\mathrm{supp}(|s_2|) \subseteq \mathrm{supp}(|t_\infty|)$. Now since $s_1, s_2$ are proper series in $K^{\mathrm{rat}}\langle\!\langle \Sigma^* \rangle\!\rangle$, there exist simple terms $t_1, t_2$ with $|t_i| = s_i$, $i = 1, 2$. By Theorem 8.5, $t_0$ and $t_1 + t_2$ are equivalent. Thus, by Lemma 11.8, $t$ is equivalent to $t_c + t_1 + \infty t_\infty$. □

We are now in the position to complete the proof of Theorem 11.1.

*Proof of Theorem 11.1, completed.* Suppose that $s, t$ are terms in the variables $x_1, \ldots, x_n$ with $|s| = |t|$. We want to show that $s$ is equivalent to $t$. Write $s = s_c + s_0 + \infty s_\infty$ and $t = t_c + t_0 + t_\infty$ as in Lemma 11.9. Since $|s| = |t|$, it follows that $|s_c| = |t_c|$, $|s_0| = |t_0|$ and $|\infty s_\infty| = |\infty t_\infty|$. Since $|s_c| = |t_c|$, $s_c$ and $t_c$ are equivalent by Lemma 11.4. Since $|s_0| = |t_0|$, $s_0$ and $t_0$ are equivalent by Theorem 8.5. Finally, since $|\infty s_\infty| = |\infty t_\infty|$, also $\mathrm{supp}(|\infty s_\infty|) = \mathrm{supp}(|\infty t_\infty|)$, which means that $\infty s_\infty$ and $\infty t_\infty$ evaluate to equal series in $\mathbb{B}^{\mathrm{rat}}\langle\!\langle \Sigma^* \rangle\!\rangle$. By Lemma 11.2, this yields that $\infty s_\infty = \infty t_\infty$ holds in $\mathcal{W}$, so that $\infty s_\infty$ and $\infty t_\infty$ are equivalent by Corollary 10.8. □

**Acknowledgement** We would like to thank the referee for helpful suggestions.